\documentclass[preprint,number,sort&compress,twocolumn,3p]{elsarticle}
\pdfoutput=1
\usepackage{rotating}
\usepackage{graphicx}
\usepackage{array}
\usepackage{amsmath}
\usepackage{color}
\usepackage{amssymb}
\usepackage{slashed}
\usepackage[pdftex,hyperfootnotes=false,pdfpagelabels]{hyperref}
\usepackage[normalem]{ulem}

\begin{document}

\title{On the LHC sensitivity for non-thermalised hidden sectors}

\tnotetext[t1]{This article is registered under preprint number: TTK-18-03}

\author{Felix Kahlhoefer}
\ead{kahlhoefer@physik.rwth-aachen.de}

\address{Institute for Theoretical Particle Physics and Cosmology (TTK), RWTH Aachen University, D-52056 Aachen, Germany}

\begin{abstract}
We show under rather general assumptions that hidden sectors that never reach thermal equilibrium in the early Universe are also inaccessible for the LHC. In other words, any particle that can be produced at the LHC must either have been in thermal equilibrium with the Standard Model at some point or must be produced via the decays of another hidden sector particle that has been in thermal equilibrium. To reach this conclusion, we parametrise the cross section connecting the Standard Model to the hidden sector in a very general way and use methods from linear programming to calculate the largest possible number of LHC events compatible with the requirement of non-thermalisation. We find that even the HL-LHC cannot possibly produce more than a few events with energy above 10 GeV involving states from a non-thermalised hidden sector.
\end{abstract}

\maketitle

\section{Introduction}

One of the central motivations for the LHC has been to probe models of dark matter (DM) in which the DM candidate is a weakly interacting massive particle (WIMP), meaning that it has mass and interaction strength comparable to other known particles (i.e.\ comparable to the electroweak scale). In such models the DM particle is predicted to be in thermal equilibrium with the bath of Standard Model (SM) particles, so that its relic abundance can be calculated via the freeze-out mechanism~\cite{Lee:1977ua}.

The non-observation of new physics at the LHC has however cast doubt on the idea of WIMP DM and has led to increased interest in alternative ideas. A simple modification of the WIMP framework is to assume that even though DM was in thermal equilibrium in the early Universe, the annihilation processes that are responsible for setting its relic abundance are \emph{secluded} from the SM and therefore much more difficult to observe at the LHC~\cite{Pospelov:2007mp}. Within this framework (recently coined ``WIMP next door''~\cite{Evans:2017kti}) it is easily possible to reconcile LHC constraints with the idea of thermal freeze-out~\cite{Duerr:2016tmh}.

Nevertheless, LHC results may also be taken to indicate that DM was not produced via thermal freeze-out and that in fact the DM particle was never in thermal equilibrium with the SM thermal bath. Among the many possible alternatives (collectively referred to as FIMPs for feebly-interacting massive particles~\cite{Bernal:2017kxu}) are models in which DM is produced via out-of-equilibrium decays (so-called superWIMPs~\cite{Feng:2003uy}) or via the freeze-in mechanism~\cite{Hall:2009bx}. A question of general interest is therefore whether the LHC may be able to also shed light on models in which the dark sector is so weakly coupled to the SM that it was never in thermal equilibrium. 

In this letter we address this question by comparing interaction rates at the LHC and in the early Universe. This comparison is based on the following simple observations:
\begin{enumerate}
 \item If the same physics can be used to describe the early Universe and the LHC, any process that can be induced at the LHC also took place in the early Universe (provided the process happens on short enough timescales). Given a temperature of the SM thermal bath, it is straight-forward to calculate the rate at which this process must have occurred.
 \item Assuming that SM states dominate the energy density of the early Universe, we can directly compare the rate of any such process (at a given temperature) to the corresponding Hubble rate. If it exceeds the Hubble rate, the process happened sufficiently frequently that the final state particles were produced with large abundance. This typically implies that the inverse process must also have occurred frequently, leading to thermal equilibrium between initial and final states.
\end{enumerate}

It is straight-forward to perform such a comparison in the context of specific models. Here we demonstrate that it is in fact also possible to compare rates at the LHC and in the early Universe in a model-independent way. To do so, we evaluate the thermalisation condition between the SM and a hidden sector\footnote{We use the term hidden sector to refer to all particles beyond the SM that are either stable or sufficiently long-lived to appear stable in the early Universe, as well as any other states that decay dominantly into these new particles.}  for a very general functional form of the cross section of the process linking the two sectors. Requiring that the two sectors do not thermalise then leads to constraints on this cross section, which can be translated into an upper bound on the rate of the same process at the LHC. This approach allows us to show that any process that does not reach thermal equilibrium in the early Universe can only induce a negligibly small number of events with sufficiently high energy at the LHC.

We conclude that, provided the Universe once was at a temperature comparable to LHC energies, any hidden sector that we could hope to observe at the LHC must have been at least partially in thermal equilibrium in the early Universe. Conversely, any hidden sector that was not in thermal equilibrium in the early Universe is unobservable for the LHC.

This letter is structured as follows. In section~\ref{sec:main} we present the derivation of the upper bound on the number of observable LHC events, first for a rather specific assumption on the functional form of the cross section and then using a general parametrisation. A detailed discussion of the underlying assumptions is provided in section~\ref{sec:discussion}. \ref{app:s-channel} takes a closer look at the case of $s$-channel resonances.

\section{Thermal equilibrium and LHC predictions}
\label{sec:main}

We consider a process that converts two SM particles into one or more hidden sector states. Rather than specifying the details of the underlying interactions, we simply assume that the total cross section $\sigma$ of this process depends in some way on the centre-of-mass energy $\sqrt{s}$ of the colliding particles. The fundamental requirement we impose is that the cross section is sufficiently small that the hidden sector states never reach thermal equilibrium (in short, we require \emph{non-thermalisation}). To make this statement quantitative, we define the reaction rate $\Gamma$ as the product of the thermally averaged cross section times relative velocity and the equilibrium number density of the SM particles in the initial state~\cite{Gondolo:1990dk}:
\begin{equation}
 \Gamma \equiv \langle \sigma v \rangle \, n^\text{eq} = \int \frac{N_c \, s^2 \, K_1 (\sqrt{s}/T)}{4 \pi^2 \, T^2} \sigma(\sqrt{s}) \, \mathrm{d}\sqrt{s} \; ,
 \label{eq:reaction}
\end{equation}
where $N_c$ denotes the number of colour degrees of freedom in the initial state and we have assumed the masses of the SM particles to be negligible.\footnote{We note that for temperatures much larger than the masses of the particles in the initial state, our implicit assumption of a Maxwell-Boltzmann distribution for $n^\text{eq}$ is not necessarily a good approximation. However, we do not want to limit ourselves to the case of either only fermionic or only bosonic initial states and therefore proceed with the simpler expression, noting that a more accurate treatment may change our results by up to a factor of two~\cite{Evans:2017kti,Belanger:2018ccd}.} The requirement that no thermal equilibrium is achieved can then be rephrased as
\begin{equation}
 \Gamma(T) < H(T) = \sqrt{\frac{4 \pi^3 \, g_\ast}{45}} \frac{T^2}{M_\text{pl}}
 \label{eq:gamma}
\end{equation}
for all temperatures $T$, where $g_\ast$ counts the number of effective relativistic degrees of freedom at temperature $T$ and $M_\text{pl}$ denotes the Planck mass.

It will be convenient to rewrite this requirement by introducing the dimensionless reaction rate
\begin{align}
 \gamma(T) & \equiv \frac{\Gamma(T)}{H(T)} \nonumber \\ & = \int \mathrm{d}\sqrt{s} \sqrt{\frac{45}{\pi \, g_\ast}} \frac{N_c \, M_\text{pl} \, s^2 \, K_1 (\sqrt{s}/T)}{8 \pi^3 \, T^4} \sigma(\sqrt{s})
\end{align}
The non-thermalisation constraint then simply becomes $\gamma(T)  < 1$ for all $T$.

Let us now consider the same process at the LHC. Clearly, the process will only be relevant if both of the particles in the initial state can be supplied via proton-proton collisions.\footnote{Hidden sector states could also be radiated from a final state, but the corresponding cross sections will be much smaller.} In this case, the leading-order production cross section at the LHC is given by
\begin{equation}
 \sigma_\text{LHC} = \int \mathrm{d}x_1 \, \mathrm{d}x_2 \, f_1(x_1) \, f_2(x_2) \, \sigma(\sqrt{s_\text{tot} \, x_1 \, x_2}) \; ,
 \label{eq:sigmaLHC}
\end{equation}
where $\sqrt{s_\text{tot}}$ is the total centre-of-mass energy of the LHC, $x_i$ denote the fraction of the proton momentum carried by the particles in the initial state and $f_i$ denote the parton distribution functions (pdfs). Here we use the \texttt{MSTW 2008 NNLO} pdfs~\cite{Martin:2009iq}, setting the factorisation scale to the partonic centre-of-mass energy $\mu_\mathrm{F} = E_\text{cm} \equiv \sqrt{s_\text{tot} \, x_1 \, x_2}$.

The total number of production processes of hidden sector states that will occur at the LHC is then given by $N_\text{LHC} = \sigma_\text{LHC} \, \mathcal{L}$, where $\mathcal{L}$ denotes the integrated luminosity. Again, it will be convenient to rewrite this expression slightly:
\begin{equation}
 N_\text{LHC} = \int \mathrm{d}\sqrt{s} \, \frac{\mathrm{d}x}{x} f_1(x) \, f_2\left(\tfrac{s}{s_\text{tot} \, x}\right) \frac{2 \, \mathcal{L} \, \sqrt{s}}{s_\text{tot}} \sigma(\sqrt{s}) \; .
 \label{eq:NLHC}
\end{equation}

\subsection{A first example}

Our aim is to find the functional form $\sigma(\sqrt{s})$ that maximises $N_\text{LHC}$ while satisfying the constraint $\gamma(T) < 1$. To gain some intuition on the nature of these requirements, let us first make a very simple ansatz and write
\begin{equation}
 \sigma(\sqrt{s}) = \sigma_0 \sqrt{s_0} \, \delta(\sqrt{s} - \sqrt{s_0}) \; ,
\end{equation}
where $\delta(x)$ is the Dirac $\delta$-function.
Such a cross section could arise for example from the exchange of an on-shell $s$-channel mediator connecting the SM initial state to the hidden sector (see~\ref{app:s-channel} for further details). With this ansatz we find
\begin{align}
 \gamma(T) & = \sqrt{\frac{45}{\pi \, g_\ast}} \frac{N_c \, M_\text{pl} \, \sqrt{s_0}^5 \, K_1 (\sqrt{s_0}/T)}{8 \pi^3 \, T^4} \sigma_0 \nonumber \\ 
 & = \sqrt{\frac{45}{\pi \, g_\ast}} \frac{N_c \, \sqrt{s_0} \, M_\text{pl} \, x_0^4 \, K_1(x_0)}{8 \pi^3} \sigma_0 \; ,
\end{align}
where in the second line we have defined $x_0 = \sqrt{s_0} / T$. It is straight-forward to see that if we vary $T$ (or equivalently $x_0$), the expression is maximised for $x_0 \approx 3.414$. Thus, the constraint $\gamma(T) < 1$ is satisfied for all $T$ exactly if
\begin{equation}
 \sigma_0 < 6.45 \cdot 10^{-6} \, \mathrm{fb} \left(\frac{1\,\mathrm{GeV}}{\sqrt{s_0}}\right) \frac{1}{N_c} \left(\frac{g_\ast(T_0)}{106.75}\right)^{1/2}\; ,
\end{equation}
where $T_0 = \sqrt{s_0} / 3.414$. 

Clearly, the non-thermalisation constraint requires the cross section connecting SM states to the hidden sector to be extremely small. We observe in particular that the constraint becomes more stringent with increasing $\sqrt{s_0}$. The reason is that in this case the non-thermalisation constraint becomes sensitive to higher energies, corresponding to earlier times and hence larger densities in the early Universe.

\begin{figure}[t]
\centering
\includegraphics[width=0.95\columnwidth]{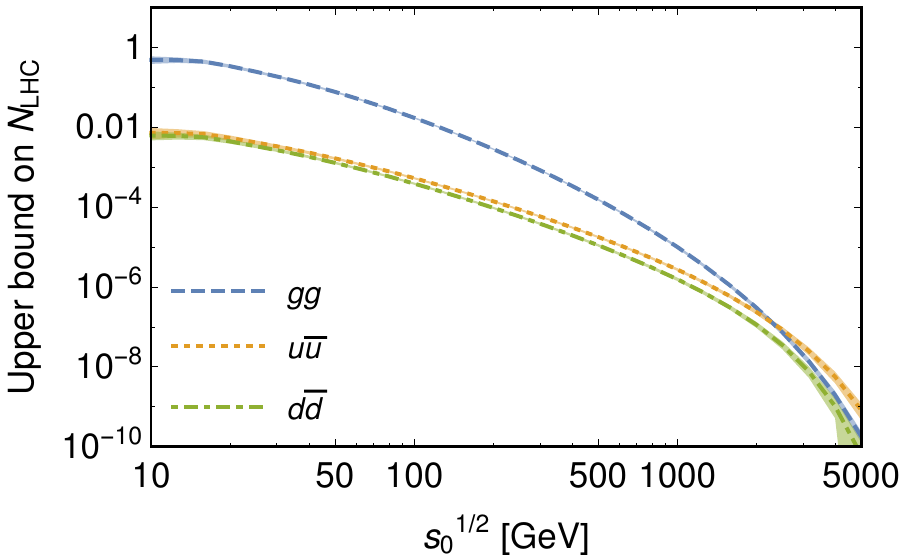}
\caption{\label{fig:ansatz} Upper bound on the number of events at the HL-LHC for a cross section of the form $\sigma(\sqrt{s}) = \sigma_0 \sqrt{s_0} \, \delta(\sqrt{s} - \sqrt{s_0})$ for different initial states. For each value of $\sqrt{s_0}$ the coefficient $\sigma_0$ is fixed in such a way that the cross section saturates the non-thermalisation constraint. Shaded bands indicated pdf uncertainties at the 90\% confidence level.}
\end{figure}

We can make use of the upper bound on $\sigma_0$ to substitute our ansatz into eq.~(\ref{eq:NLHC}) and calculate the maximum number of events that can be expected at the LHC for a cross section of this particular form. The result is shown in figure~\ref{fig:ansatz} for an integrated luminosity of $\mathcal{L} = 3000\,\mathrm{fb^{-1}}$ for three different combinations of initial states: $gg$, $u\bar{u}$ and $d\bar{d}$. We make the following observations:
\begin{enumerate}
 \item The predicted number of events depends sensitively on $\sqrt{s_0}$. This is a direct consequence both of the non-thermalisation constraint becoming weaker for smaller $\sqrt{s_0}$ and of the pdfs preferring smaller partonic centre-of-mass energies.
 \item We find significantly larger values of $N_\text{LHC}$ for the case of a $gg$ initial state than for a $u\bar{u}$ initial state, again a direct consequence of the larger gluon pdfs at small momentum fraction.
 \item In any case, the upper bounds on $N_\text{LHC}$ are extremely tight. Even for $\sqrt{s_0} = 10\;\mathrm{GeV}$ and a $gg$ initial state, we find $N_\text{LHC} < 0.49$, making a discovery of this process impossible.\footnote{In principle, even a single event could allow for a discovery, if the signature is so distinctive that it cannot be missed or confused with SM backgrounds. Since we do not specify the details of the hidden sector, we cannot entirely exclude this possibility although it seems rather unlikely.}
\end{enumerate}

In principle, the bound on $N_\text{LHC}$ could be further relaxed by going to even smaller values of $\sqrt{s_0}$. However, at this point our description becomes increasingly questionable, because non-perturbative effects become important for the description of the initial state. Moreover, processes with such small partonic centre-of-mass energy are very challenging to observe at the LHC and would be more suitable for searches at low-energy colliders like Belle II. To avoid these complications, we will not consider values of $\sqrt{s_0}$ below 10 GeV. In fact, even for $\sqrt{s_0} = 10 \, \text{GeV}$ pdf uncertainties are already substantial. We estimate these uncertainties by varying the pdfs around the central set following the procedure described in ref.~\cite{Martin:2009iq}. The shaded bands in figure~\ref{fig:ansatz} illustrate the resulting changes in $N_\text{LHC}$ at the 90\% confidence level. For $\sqrt{s_0} = 10\,\text{GeV}$ the uncertainty amounts to approximately 20--30\%.

We conclude that the simple ansatz chosen above leads to a very stringent bound on $N_\text{LHC}$. Nevertheless, it is far from clear that this ansatz comes close to the optimal functional form of the cross section. After all, we found that $\gamma(T) \approx 1$ only for a very small range of temperatures. We therefore expect that it should be possible to relax the bound on $N_\text{LHC}$ by considering more general forms of $\sigma(\sqrt{s})$. 

\subsection{General cross sections}

Since it is difficult to consider completely arbitrary variations of $\sigma(\sqrt{s})$, we will instead consider a discretised version of the problem (see ref.~\cite{Feldstein:2014gza} for a similar approach in the context of DM direct detection experiments). For this purpose, we define a (potentially large) number of discrete values for $\sqrt{s}$:
\begin{equation}
 \sqrt{s_i} = \sqrt{s_0} \Delta^i \; ,
\end{equation}
where $i = 0, \ldots, m$ and $\Delta > 1$ defines the step size. Since we are only interested in quantities involving the integral of $\sigma(\sqrt{s})$, we can then approximate the cross section by
\begin{equation}
 \sigma(\sqrt{s}) \approx \sum_i \sigma_i \sqrt{s_i} \, (\Delta - 1) \, \delta(\sqrt{s} - \sqrt{s_i})
 \label{eq:para}
\end{equation}
with $\sigma_i = \sigma(\sqrt{s_i})$. This approximation can be made for any cross section that does not vary too rapidly on each interval $[\sqrt{s_i}, \sqrt{s_{i+1}}]$, and it becomes exact in the limit $\Delta \to 1$. Rather than considering cross sections of arbitrary functional form, we can therefore simply study cross sections of the form of eq.~(\ref{eq:para}) with arbitrary coefficients $\sigma_i$. As we will see below, writing $\sigma(\sqrt{s})$ as a sum of $\delta$-functions (rather than e.g.\ as a piecewise constant function) has the advantage that it leads to a significant simplification of the relevant equations.

Making use of the approximation introduced in eq.~(\ref{eq:para}), the number of predicted events at the LHC can be written as
\begin{equation}
  N_\text{LHC} = \sum_i b_i \, \sigma_i
\end{equation}
with
\begin{equation}
  b_i =  (\Delta - 1) \int \frac{\mathrm{d}x}{x} f_1(x) \, f_2\left(\tfrac{s_i}{s_\text{tot} \, x}\right) \frac{2 \, \mathcal{L} \, s_i}{s_\text{tot}}  \; .
\end{equation}

The second important simplification that we make is to assume that it is sufficient to check the non-thermalisation constraint only for discrete values of $T$. Specifically, we define
\begin{equation}
 T_j = T_0 \Delta^j \;
\end{equation}
for $j = 1,\ldots,n$ and appropriate values $T_0 < \sqrt{s_0}$ and $n > m$, such that all relevant temperatures are covered. With this definition, we find that $\gamma_j \equiv \gamma(T_j)$ can be written as
\begin{equation}
 \gamma_j = \sum_i a_{ji} \sigma_i
\end{equation}
with
\begin{equation}
 a_{ij} = (\Delta - 1) \sqrt{\frac{45}{\pi \, g_\ast}} \frac{N_c \, M_\text{pl} \, \sqrt{s_i}^5 \, K_1 (\sqrt{s_i}/T_j)}{8 \pi^3 \, T_j^4} \; .
\end{equation}

We have therefore reduced the problem to a well-known problem of linear programming: the maximisation of $\mathbf{b} \cdot \mathbf{x}$ with respect to a vector $\mathbf{x} = (\sigma_i)$ with positive entries for a given vector $\mathbf{b}$, subject to a set of constraints given by $\mathbf{A} \cdot \mathbf{x} \leq \mathbf{1}$. This optimisation problem can be easily solved with well-known numerical methods. For sufficiently small step size $\Delta$ and sufficiently large $m$ and $n$, the value of the maximum should become independent of the precise value of $\Delta$ and of the choice of $T_0$. We find that this is indeed the case, and that choosing $\Delta = 1.1$ (corresponding to $m, n \sim 100$) is fully sufficient.

\subsection{Results}

For concreteness, let us consider the case of a $gg$ initial state and set $\sqrt{s_0} = 10\,\text{GeV}$ and $T_0 = 1\,\text{Gev}$. For $\sqrt{s_\text{tot}} = 13\,\mathrm{TeV}$ and $\mathcal{L} = 3000\,\mathrm{fb^{-1}}$ we find $N_\text{LHC} < 0.62$. This value should be compared with the value $N_\text{LHC} < 0.49$ that we found for the much simpler ansatz above. In fact, the similarity of the two results is no coincidence, as will become clear by inspecting the optimum solution.

\begin{figure*}[t]
\centering
\includegraphics[width=0.93\columnwidth]{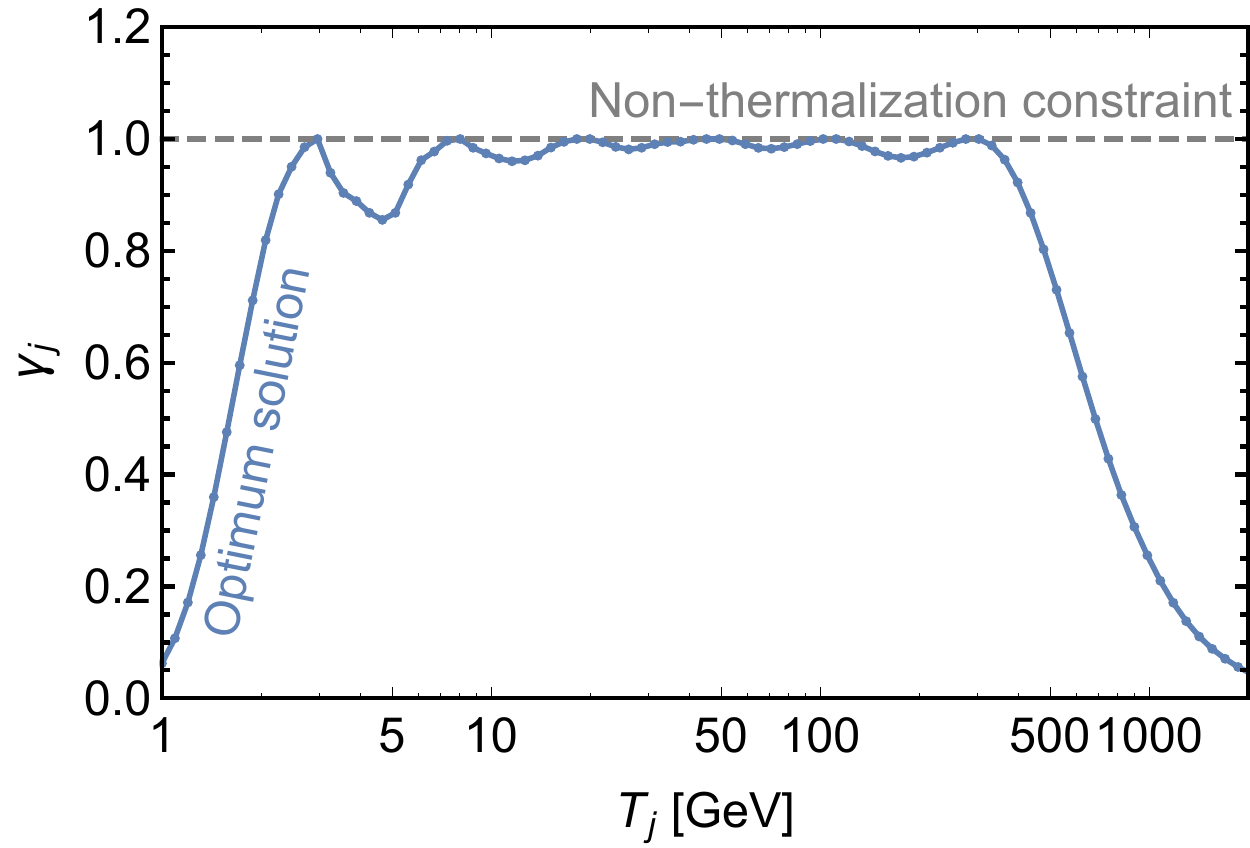}\qquad
\includegraphics[width=0.965\columnwidth]{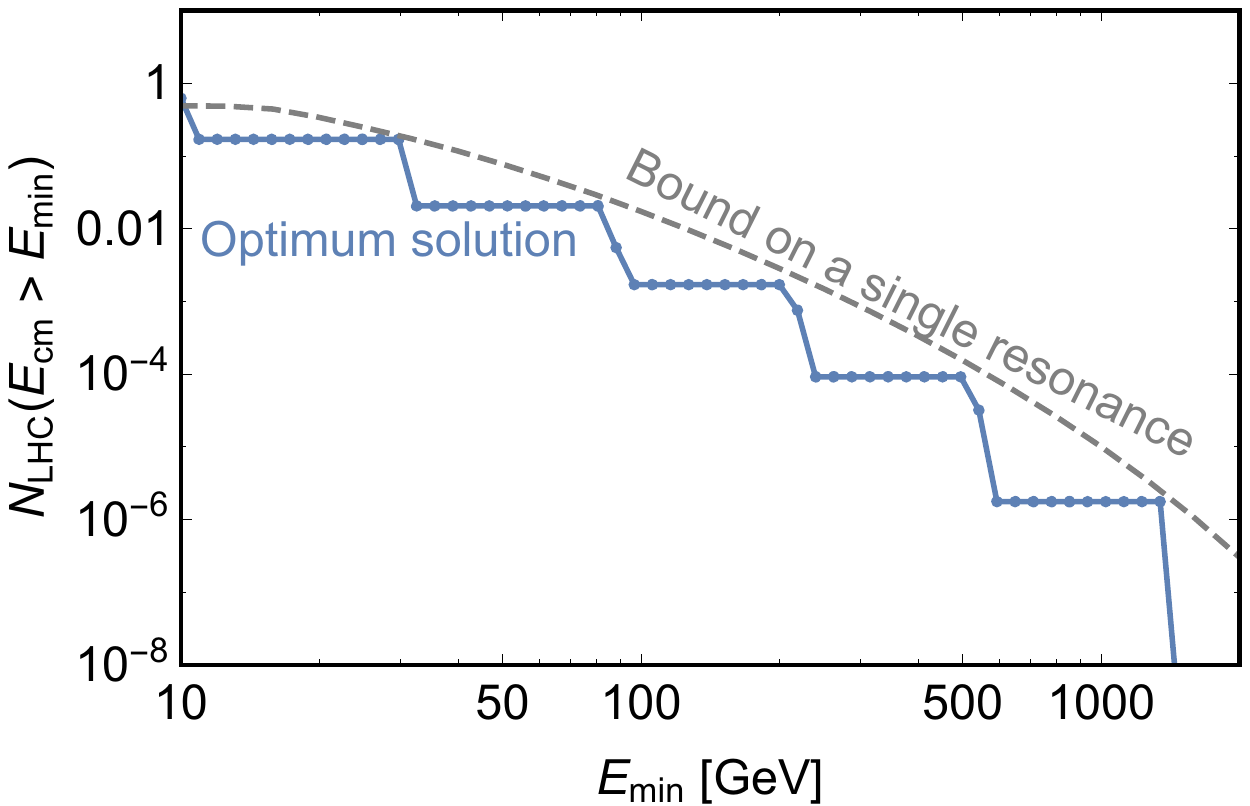}
\caption{\label{fig:solution} Illustration of the optimum cross section, which satisfies the non-thermalisation constraint for all temperatures (visualised in the left panel) while maximizing the number of predicted events at the LHC (visualised in the right panel). This example considers a $gg$ initial state with an integrated luminosity of $\mathcal{L} = 3000\,\mathrm{fb^{-1}}$ at $\sqrt{s_\text{tot}} = 13\,\text{TeV}$. The low-energy cut-off has been set to $\sqrt{s_0} = 10\,\mathrm{GeV}$. For easier visualisation we show continuous lines connecting the discrete values used for the optimisation. The dashed lines indicate the constraint $\gamma_i < 1$ in the left panel and the bound on $N_\text{LHC}$ obtained for a single $\delta$-function with $\sqrt{s_0} = E_\text{min}$ in the right panel (see figure~\ref{fig:ansatz}).}
\end{figure*}

The central observation is that the optimum solution has $\sigma_i = 0$ for almost all values of $i$. The few non-zero values of $\sigma_i$ are well-separated and each (nearly) saturate the bound derived above for the case of a single $\delta$-function. In this way, the individual contributions conspire to give $\gamma(T) \approx 1$ over a wide range of temperatures. This behaviour is illustrated in the left panel of figure~\ref{fig:solution}, which shows that in this particular case the optimum solution consists of a sum of six separate $\delta$-functions.

We emphasise that the number of $\delta$-functions contributing to the optimum solution is not an artefact of the chosen step size, which is in fact much smaller than the distance between two $\delta$-functions. Instead, it is a generic feature of the optimum solution and is robust under a change of step size. In other words, the ``distance'' between the individual $\delta$-functions is optimal~-- introducing any additional contribution would strengthen the non-thermalisation constraint and hence lead to a stronger bound on $N_\text{LHC}$.

This observation makes it clear why the optimum solution is so close to the one obtained for a single $\delta$-function: The dominant contribution to $N_\text{LHC}$ stems from the $\delta$-function at $\sqrt{s_0}$, while the additional $\delta$-functions at higher energies only give sub-leading contributions. This is illustrated in the right panel of figure~\ref{fig:solution}, which shows the predicted number of LHC events with partonic centre-of-mass energy $E_\text{cm}$ above a minimum energy $E_\text{min}$ as a function of that minimum energy. The dashed grey line shows for comparison the bound on $N_\text{LHC}$ obtained for a single $\delta$-function with $\sqrt{s_0} = E_\text{min}$ (see figure~\ref{fig:ansatz}). For the specific case considered here, we find that the first three $\delta$-functions contribute $0.45$, $0.15$ and $0.019$ to the total of $0.62$ predicted events.

To conclude this discussion, we emphasise that the numbers discussed above depend of course on the somewhat arbitrary choice of $\sqrt{s_0}$. To first approximation, reducing $\sqrt{s_0}$ by a factor $r$ will relax the bound on $N_\text{LHC}$ by the same factor. However, even if we allow $\sqrt{s_0}$ as low as $1\,\mathrm{GeV}$ (and consider $\sqrt{s_\text{tot}} = 14\,\mathrm{TeV}$), we cannot obtain more than $\mathcal{O}(10)$ events with in full data set of the HL-LHC, almost all of which would be at extremely low energies and therefore likely unobservable.

\section{Discussion}
\label{sec:discussion}

We have shown in a very general way that any process connecting the SM to a hidden sector that was not in thermal equilibrium in the early Universe is unobservable at the LHC. Turning this argument around, we conclude that any hidden sector that can be produced and observed at the LHC must have been at least partially in thermal equilibrium with the bath of SM particles in the Early Universe. 

We emphasise that this does not necessarily mean that the LHC can only probe DM models in which the DM particle itself was in thermal equilibrium. For example, the DM particle may be produced from the decays of another metastable hidden sector particle, which itself was in thermal equilibrium (as in the case of gravitino DM produced in the decays of the next-to-lightest supersymmetric particle).

An immediate consequence of these observations is that it is impossible for the LHC to test most realisations of the freeze-in mechanism, in which no part of the hidden sector thermalises with the SM. More complex freeze-in models may be testable, for example if they contain a new state that couples only very weakly to the hidden sector and decays dominantly into SM particles. However, it will be very challenging for the LHC to establish a connection to the DM problem from such observations alone.

Three fundamental assumptions enter our analysis:
\begin{enumerate}
 \item We have assumed that the early Universe reached temperatures comparable to the energies accessible for the LHC. Clearly, if the reheating temperature is significantly smaller than LHC energies, the LHC may probe processes that have never been in thermal equilibrium (see ref.~\cite{Co:2015pka}).
 \item We have assumed that the physics relevant for the interactions of the hidden sector in the early Universe is the same as the physics relevant to the LHC. This assumption could potentially be violated by phase transitions in the early Universe or by thermal effects suppressing certain processes (see~\cite{Baker:2016xzo,Baker:2017zwx} for an example).
 \item We have assumed a standard cosmological history up to energies of a few TeV. A mechanism that would significantly increase the expansion rate (e.g.\ by introducing a large number of additional relativistic degrees of freedom) would relax the non-thermalisation constraint and hence our bound on the number of observable LHC events~\cite{DEramo:2017ecx}.
\end{enumerate}

While these assumptions clearly limit the generality of our results, they are essentially unavoidable when trying to connect LHC physics to early Universe cosmology in a predictive way. At the same time they provide useful guidance for the ways in which early-Universe cosmology must be modified in order to evade our conclusions. In other words, our assumptions can be seen as a list of loopholes that may be exploited to obtain observable LHC signatures from non-thermalised hidden sectors. In any case, we emphasise that no further assumptions have been made. In particular, by considering cross sections with completely arbitrary dependence on the centre-of-mass energy, we avoid the need to specify the way in which the hidden sector interacts with the SM.

With growing interest in models of FIMP DM with non-thermal production mechanisms it becomes more and more urgent to develop new experimental strategies to probe these theories. We have shown that the LHC is not sensitive to non-thermalised hidden sectors, since the interactions are either too weak or occur dominantly at too low energies to be observable. Further work is needed to establish whether low-energy colliders or beam-dump experiments are more promising~\cite{Alexander:2016aln}, or whether we need to rely on non-collider experiments, such as low-threshold direct detection experiments~\cite{Essig:2011nj,Battaglieri:2017aum}, to make further progress. The model-independent method proposed in this work may provide a useful tool for future studies of these questions.

\section*{Acknowledgements}

FK would like to thank Stefania Gori, Andreas Goudelis, Jan Heisig, Joachim Kopp, Kai Schmidt-Hoberg, Jessie Shelton, Susanne Westhoff, Sebastian Wild and Bryan Zaldivar for valuable comments on the manuscript and all participants of the workshop ``The Future of Searches for Invisible Particles'' (14--15 December 2017, Aachen, Germany) for inspiring discussions. This work is supported by the DFG Emmy Noether Grant No.\ KA 4662/1-1.

\appendix

\section{Resonant thermalisation}
\label{app:s-channel}

In this appendix we take a closer look at the case where the interactions between the SM and the hidden sector are communicated by an $s$-channel mediator, i.e.\ a state that can be resonantly produced from SM initial states both at the LHC and in the early Universe. The mediator will be in equilibrium with the thermal bath if $\Gamma_R > H(T \sim M_R)$, where $\Gamma_R$ and $M_R$ denote the total width and mass of the resonance, respectively. If, however, the partial width for decays into the hidden sector is sufficiently small $\Gamma(R \to \text{hidden sector}) < H(T \sim M_R)$, decays of the mediator may populate the hidden sector without bringing it into equilibrium with the thermal bath. In this appendix we show that if $\Gamma(R \to yy) < H(T \sim M_R)$ for a given hidden sector final state $yy$ it follows automatically that $\gamma(T \sim M_R) < 1$ for any process $xx \to R \to yy$. 

Let us consider a specific SM initial state called $xx$. First of all, we make use of the narrow-width approximation to write
\begin{equation}
 \sigma_{xx \to R \to yy} = \sigma_{xx \to R} \times \text{BR}(R \to yy) \; .
\end{equation}
The central observation is now that the cross section $\sigma(xx \to R)$ for the production process and the partial width $\Gamma(R \to xx)$ for the inverse process depend on the same matrix element and therefore differ only by phase space factors. This enables us to write
\begin{align}
 \sigma_{xx \to R \to yy} = & \pi^2 c \, \delta(\sqrt{s} - M_R) \tfrac{\Gamma(R \to xx)}{N_c \, M_R^2} \tfrac{\Gamma(R \to yy)}{\Gamma_R} \nonumber \\
 = & \pi^2 c \, \delta(\sqrt{s} - M_R) \tfrac{\Gamma(R \to yy)}{N_c \, M_R^2} \, \text{BR}(R \to xx)
 \label{eq:sigma_res}
\end{align}
where $N_c$ counts the colour degrees of freedom of $x$ and $c$ is an order unity numerical pre-factor. For a $gg$ initial state and a scalar resonance, one finds for example $c = 1 / 2$~\cite{Djouadi:2005gi}. 

We can now substitute eq.~(\ref{eq:sigma_res}) into eq.~(\ref{eq:reaction}) to calculate an upper bound on the reaction rate for the $2 \to 2$ process. For $T \sim M_R$, and making use of $\text{BR}(R \to xx) \leq 1$, we find
\begin{equation}
 \langle \sigma_{xx \to R \to yy} v\rangle \, n^\text{eq}_x \lesssim \frac{c}{4} \, \Gamma(R \to yy) \; .
\end{equation}
Hence, if $\Gamma(R \to yy) < H(T \sim M_R)$, it follows that the same is true for the left-hand side (since $c \sim \mathcal{O}(1)$). This is equivalent to
\begin{equation}
 \gamma(T \sim M_R) < 1 \; .
\end{equation}
In other words, for the case of an $s$-channel resonance the non-thermalisation constraint that we impose is necessary but not sufficient to ensure that the hidden sector does not thermalise with the SM. Imposing instead $\Gamma(R \to yy) < H(T \sim M_R)$ would lead to even stronger bounds on $N_\text{LHC}$ than what is shown in figure~\ref{fig:ansatz}.

\end{document}